# Brillouin-Mandelstam Spectroscopy of Stress-Modulated Spatially Confined Spin Waves in Ni Thin Films on Piezoelectric Heterostructures

Fariborz Kargar[1,2,*], Michael Balinskiy[1,2], Howard Chiang[1,2], Andres Chavez[3], John Nance[3], Alexander Khitun[1,2], Gregory P. Carman[3], and Alexander A. Balandin[1,2,*]

[1]Phonon Optimized Engineered Materials (POEM) Center, Department of Electrical and Computer Engineering, University of California — Riverside, Riverside, California 92521 USA

[2]Spins and Heat in Nanoscale Electronic Systems (SHINES) Center, University of California – Riverside, Riverside, California 92521 USA

[3]Department of Electrical Engineering, University of California – Los Angeles, Los Angeles, California 92521 USA

**Abstract:** We report results of micro-Brillouin-Mandelstam light scattering spectroscopy of thermal magnons in the two-phase synthetic multiferroic structure consisting of a piezoelectric $[Pb(Mg_{1/3}Nb_{2/3})O_3]_{(1-x)}$–$[PbTiO_3]_x$ (PMN-PT) substrate and a Ni thin film with the thickness of 64 nm. The experimental data reveal the first two modes of the perpendicular standing spin waves (PSSW) spatially confined across the Ni thin film. A theoretical analysis of the frequency dependence of the PSSW peaks on the external magnetic field reveals the asymmetric boundary condition, *i.e.* pinning, for variable magnetization at different surfaces of the Ni thin film. The strain field induced by applying DC voltage to PMN-PT substrate leads to a down shift of PSSW mode frequency owing to the magneto-elastic effect in Ni, and corresponding changes in the spin wave resonance conditions. The observed non-monotonic dependence of the PSSW frequency on DC voltage is related to an abrupt change of the pinning parameter at certain values of the voltage. The obtained results are important for understanding the thermal magnon spectrum in ferromagnetic films and development of the low-power spin-wave devices.

---

[*]Corresponding authors: fariborz.kargar@ucr.edu ; balandin@ece.ucr.edu ; http://balandingroup.ucr.edu/

Recently, synthetic multiferroic heterostructures, materials with simultaneous magnetic and ferroelectric ordering, have attracted significant interest owing to a possibility of engineering their electric and magnetic properties by varying the thickness ratios of the constituent materials [1–5]. Similar to single-phase multiferroics [6,7], synthetic multiferroics exhibit strong electro-magnetic coupling. For example, up to 150 degree easy axis rotation was observed in CoPd/PZT structures [3]. It has been demonstrated experimentally that the frequency of the microwave planar resonator, consisting of yttrium iron garnet (YIG) and ferroelectric barium strontium titanate (BST) thin films, can be tuned both electrically and magnetically [8]. Recent reports have also shown the spin-wave frequency modulation in a permalloy film by introducing strain in gadolinium molybdate [9] and PMN-PT [10] substrates via applications of DC voltage. Spin-wave excitation and detection by synthetic multiferroics comprised of PMN-PT/Ni/Py has also been demonstrated [2]. In many of the studied multiferroic heterostructures, a thin layer of magnetostrictive material *e.g.* nickel is deposited on a piezoelectric material, which makes it possible to control the thermal and coherent magnons via the stress-mediated coupling, induced by application of an electric field to the substrate [11]. For this reason, the interaction of the piezoelectric substrate with the thin ferromagnetic layer and their magneto-elastic coupling are of utmost importance.

While magnon spectrum of a large, *i.e.* bulk, magnetically ordered medium depends only on its material parameters and an applied magnetic field, it can be strongly modified in magnetic thin films by the boundary conditions at the interface of the film with a substrate [12–14]. The properties of thermal and coherently excited magnons, *i.e.* spin waves, in ferro☐, ferri☐, and antiferro☐magnetic films and multilayer structures have been described in numerous reports [15–23]. Brillouin-Mandelstam spectroscopy (BMS) has been recognized as an important tool for investigation of spin-wave excitations in magnetic micro- and nano-structures [24]. This technique offers unique advantages over the all-electrical approaches, *i.e.* inductive voltage measurements [25,26], by allowing the detection of extremely weak signals, including those from thermal magnons [18]. In this regard, there is a growing demand for BMS utilization in the study and development of magnonic spintronic devices [27–30]. Similar to thermal phonons, the incoherent thermal magnons are always present in magnetic materials. The latter affects the magnetization stability of the magnetic materials and structures [31]. The thermal magnons determine the equilibrium magnetization and affect thermodynamic characteristics of materials [32]. In addition, thermal magnons contribute to relaxation of the coherent magnons [33]. The perpendicular standing spin waves (PSSW) are of special interest since they also strongly influence the performance of magnonic devices [34]. These types of waves have been studied in micrometer and nanometer thick iron, nickel, and cobalt films on glass or silicon substrates [16,18,35]. In this work, we use micro-BMS (μ-BMS) to investigate PSSW in nanometer-scale thin films of nickel deposited on a piezoelectric $[Pb(Mg_{1/3}Nb_{2/3})O_3]_{(1-x)}$–$[PbTiO_3]_x$ (PMN-PT) substrate. Our μ-BMS data clearly demonstrates the spatial confinement induced modification of the thermal magnon spectra in the synthetic multiferroic structures. The ability to control the thermal magnons by applying an electric field in synthetic multiferroic structures with restricted geometry opens a new route to development and optimization of the magnonic spintronic devices. The confinement induced changes in magnon spectrum in nanometer thickness structures can also have important effects on the coherent magnon damping.

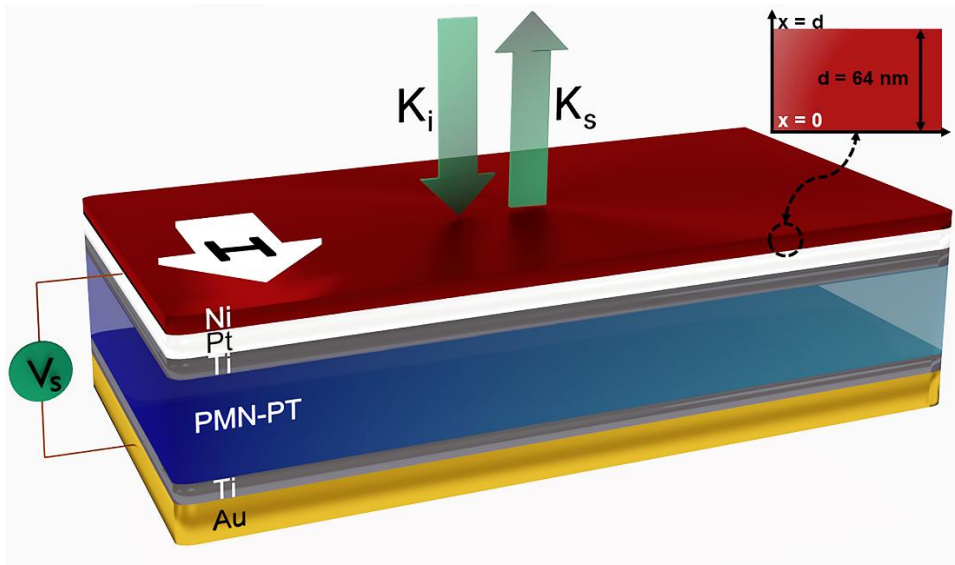

**Figure 1:** Schematic of the device structure showing the thin layer of polycrystalline Ni deposited on a PMN-PT substrate. The circle shows the magnified section of the Ni layer. The green arrows show the incident and scattered light in μ-BMS experiments in the backscattering configuration.

The cross-section of a test structure used in this study is shown in Figure 1. The sample was fabricated on a 10 mm × 10 mm × 0.5 mm single crystal (011) PMN-PT substrate (TRS Technologies). From the bottom to the top, it consists of the following layers: 30 nm Au, 5 nm Ti, 0.5 mm PMN-PT (011 single crystal cut), 5 nm Ti, 30 nm Pt, and 64 nm Ni. Prior to all metal depositions, the sample is cleaned with acetone, methanol, IPA, followed by a 500 W $O_2$ plasma clean. Ti is chosen as an adhesion layer for the Au and Pt electrodes. The electrodes are deposited in such a way that the sample can be poled prior to deposition of the magnetic layers. A custom brass holder and a voltage amplifier are used to apply an electric field to the sample for poling. The applied electric field is linearly ramped from 0 MV/m to 0.8 MV/m over a one-minute period and is then held constant for the same amount of time. Following this procedure, the field is removed at the same rate as it was applied. The latter is important because poling the sample after deposition of the magnetic layers can lead to residual stresses that cannot be overcome with the voltage-induced piezoelectric stresses. Pt is used because of its high electrical conductivity and superior resistance to oxidation. The magnetic properties of the Ni film were characterized with the magneto-optic Kerr effect (MOKE). A typical easy-plane hysteresis curve obtained for the perpendicular magnetized film is shown in Figure 2. The magnetization saturation of the Ni film has been determined to be $4\pi M_0 = 2640$ Oe, which is lower than the reported bulk values ~6000 Oe [20]. Thin films of Ni are known to have lower saturation magnetization than bulk samples [36]. The lower magnetization of the thin films can be attributed to partial surface oxidation or fabrication processes [36].

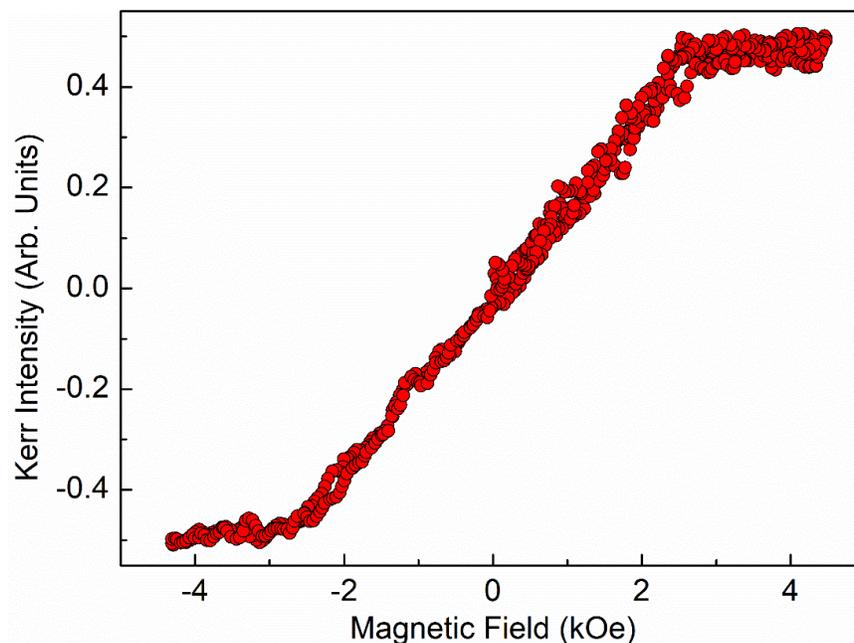

**Figure 2:** Easy-plane hysteresis curve for the perpendicularly magnetized Ni thin film as a function of an external magnetic field.

We used μ-BMS to monitor thermal magnons in a backscattering configuration. The light source was a single frequency solid-state diode laser operating at $\lambda = 532$ nm. The laser light was focused normal to the sample by using a 100× objective lens with a large numerical aperture. The laser spot size with diameter of ~1-2 μm was located at the center of the sample in order to avoid any edge effects. In addition, the average laser power at the sample's surface was kept less than 2 mW during the experiment in order to avoid overheating the Ni film. The scattered light was collected with the same lens for 15 minutes and directed to the high resolution six-pass (3+3) tandem Fabry-Perot interferometer (JRS Instruments). The incident light was linearly *p*-polarized. Since in the ferromagnetic materials, spin waves rotate the polarization of the incident photons by 90 degrees [10] a polarizer in the scattered light path has been used to allow for transmission of only the *s*-polarized light into the interferometer. Details of our BMS experimental setup and measurement procedures have been reported elsewhere in the context of different material systems [37,38]. In the present study, the bias magnetic field $H_0$ was applied in-plane with the Ni film surface and perpendicular to the light scattering plane. Taking into account a relatively small penetration length of light in metals, which is much smaller than the wavelength of the excitation laser light (~12 nm at $\lambda = 532$ nm for Ni [39]), and the uncertainty principle, the conservation of the normal component of the $k$-vector of light was not satisfied in the considered scattering processes. The latter means that thermal magnons with different frequencies and wave vectors can participate in the scattering processes and contribute to the accumulated BMS spectra [18,40]. More detailed explanation is provided below.

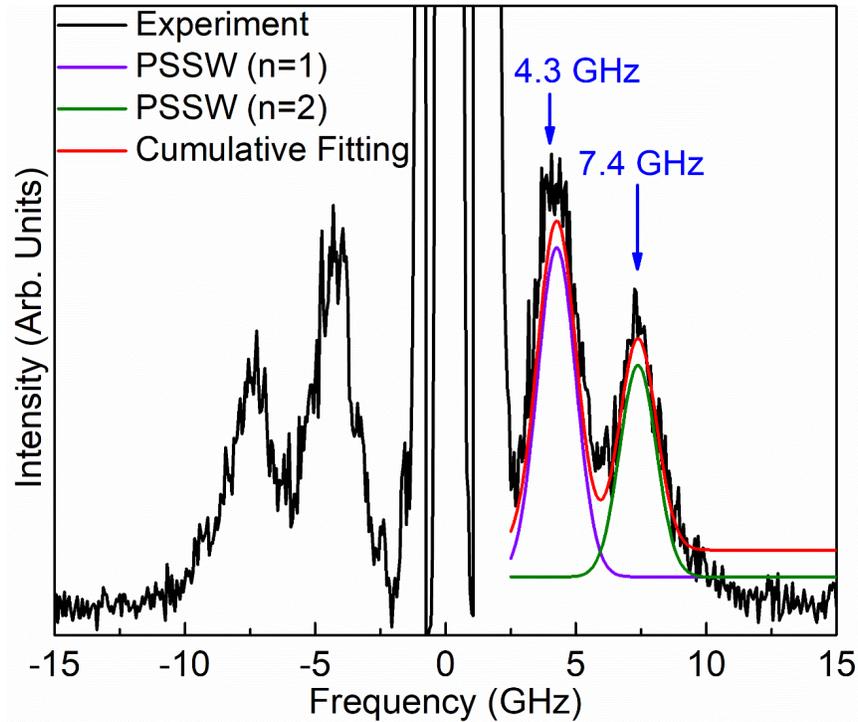

**Figure 3:** BMS spectrum of Ni thin film measured at an external magnetic field of 440 Oe. The peaks at 4.3 GHz and 7.4 GHz are attributed to the first (n=1) and second (n=2) PSSW modes spatially confined across the film thickness. The violet and green curves are individual Lorentzian fittings of each peak whereas the red curve shows the cumulative fitting to the experimental data.

A representative BMS spectrum of Ni thin film on PMN-PT substrate is shown in Figure 3. The data are collected at the external magnetic field of $H_0 = 440$ Oe. As can be seen, there are four distinct peaks at ±4.3 GHz and ±7.4 GHz frequencies. The peaks with the positive and negative frequency shifts correspond to the anti-Stokes and Stokes processes, respectively. The peaks were accurately fitted with individual Lorentzian functions (violet and green curves) to define the spectrum position of each maximum. The red curve is a cumulative fitting of all individual peaks, which matches perfectly with the experimental BMS data. As it was mentioned above, owing to the small penetration length of the light into the metal, only the in-plane component of the light wave vector is conserved during the light scattering process [18,40]. In our case of the tangentially magnetized thin film, in addition to the PSSW modes, one also expects surface magnon modes, which are propagating perpendicular to the bias magnetic field $H_0$ and in the in-plane surface of the ferromagnetic film. These modes are referred to as Damon-Eshbach (D-E) modes [40,41]. Since in our experiment the incident light is normal to the sample's surface, one would expect to observe a peak, corresponding to the D-E mode with $q_\parallel = 0$, where $q_\parallel$ is the in-plane component of the incident light wave vector. However, in the ferromagnetic thin films, the D-E mode with $q_\parallel = 0$ is the ferromagnetic resonance (FMR) mode with a uniform distribution of the variable magnetization **m** through the thickness and in-plane direction of the film. This case corresponds to unpinning boundary conditions on both surfaces of thin films [42,43]. One should note that because of the limited numerical aperture of the focusing lens, all D-E modes with larger in-plane wave-vectors contribute to the light scattering, which result in a continuous background spectrum (see Figure 3). Due to the uncertainty principal, in the direction perpendicular to the surface of the film, all PSSW modes with the wave vectors $q$ satisfying $k - \delta k \leq q \leq k + \delta k$ also contribute to the light scattering. Here $k = 2n_1 k_0 = 4\pi n_1/\lambda$ ($k_0 = 2\pi/\lambda$ is the wavevector magnitude of light) is the normal-to-surface component of the wave vector of the excited quasiparticle (phonon, magnon) inside the medium as a result of interaction with the light in the backscattering configuration. The uncertainty in the wave vector is defined as $\delta k/k \sim 2n_2/n_1$ where $n_1$ and $n_2$ are the real and imaginary parts of the refractive index, respectively [18,40]. For Ni, $n_1 = 1.8775$ and $n_2 = 3.4946$ at $\lambda = 532$ nm [39] and therefore, all PSSW modes with $0 \lesssim q \lesssim 0.2094$ nm$^{-1}$ should be present in the BMS spectrum. Based on this, we assign the observed peaks in Figure 3 as the first (n=1) and second (n=2) PSSW modes. Similar thermal magnon peaks were observed previously in 60 nm thick Py film deposited on glass substrate [44]. Below we provide additional arguments supporting our peak assignment.

In order to further elucidate the nature of the observed peaks, we carried out a number of BMS measurements at different bias magnetic field $H_0$. In Figure 4, we present a summary of the sequential BMS measurements, showing the position of the peaks as a function of the external magnetic field $H_0$ changing from 100 Oe to 960 Oe. The data reveal almost synchronous shift to higher frequencies for both modes. The frequency difference between the two distinctive peaks is ~3.4 GHz at 300 Oe and ~3.2 GHz at 900 Oe. In order to explain the observed phenomena, we start with the general dispersion relation for spin waves as described in Ref. [42]:

$$f_q^2 = \gamma^2 \cdot (H_{eff} + Dq^2) \cdot (H_{eff} + Dq^2 + 4\pi M_0). \tag{1}$$

In this equation, $f_q$ is the frequency of the spin waves, $\gamma = 3$ MHz/Oe is the gyromagnetic ratio [20], $D = 3.1 \times 10^{-9}$ Oe.cm$^2$ is the exchange constant, $H_{eff}$ is the effective magnetic field, which is the sum of the external magnetic field $H_0$ and the anisotropy magnetic field $H_a$, $q$ is the wave-vector of PSSW mode, and $4\pi M_0 = 2640$ G$^{-1}$ is the saturation magnetization of the film, respectively. The wave vectors of the spatially confined PSSW modes, $q_n$, are defined from the Landau–Lifshitz–Gilbert equation for the precession motion of the magnetization $m$ with the following boundary conditions [42,43]:

$$(\xi_1 m + dm/dx)_{x=0} = 0 \qquad (2)$$
$$(\xi_2 m + dm/dx)_{x=d} = 0.$$

Here $m$ is the variable magnetization, while $\xi_1$ and $\xi_2$ are the pinning parameters at the corresponding boundary surfaces of $x = 0$ (Ni-substrate interface) and $x = d$ (Ni-air interface) of the Ni film of the thickness $d$, respectively. The pining parameters can differ significantly owing to the difference in the Ni-air and Ni-substrate surfaces. In order to fit the experimental data, we used the asymmetric boundary conditions $\xi_1 > 0$ and $\xi_2 = 0$, respectively. The specific values of $q_n$ for different PSSW modes can be found from the following equation and then substituted in the dispersion relation of Eq. (1) in order to define the frequency of the corresponding PSSW mode [42,43]:

$$\cot(q_n d) = q_n d / 2\xi \qquad (3)$$

As a result of the iterative calculations, at $\xi_1 = 0.918$, the red and blue dashed fitting lines were obtained along with the values $q_1 d = 1.05$ and $q_2 d = 3.61$ (see Figure 4). The calculated curves are in excellent agreement with the experimental data. To understand the pinning effect better, we also examined two extremes cases of the totally symmetric and asymmetric boundary conditions at two interfaces. In case of symmetrical boundary conditions where $\xi_1 = \xi_2 = \xi$ [42,43], the wave vectors of PSSW modes are governed by the condition $q_n = (2n - 1)\pi$, $n \in N = 1,2,...$, which does not depend on the pinning parameter $\xi$. In this case, for the first two PSSW modes, one would obtain $q_1 d = \pi$ and $q_2 d = 3\pi$, respectively. Substitution these values to Eq. (1), we obtain the frequencies for the first two PSSW modes, which differ significantly from the experimentally measured values. On the other hand, considering the asymmetrical case where the pinning parameter $\xi_1 \to \infty$ (total pinning at Ni-substrate interface) and $\xi_2 = 0$ (no pinning at Ni-air interface), one would obtain the conditions $q_1 d = \pi/2$ and $q_2 d = 3\pi/2$ [42,43], which again results in frequencies considerably different from the experimental values. Assuming different values for pinning parameter at dissimilar interfaces of Ni-air and Ni-substrate means that most probably we have a case of asymmetrical pinning with unpinned spins at the free surface of Ni film and partial pinning spins at Ni-substrate interface. Given this, one should expect that the pinning parameter $\xi_1$ at the Ni-substrate interface should change when the Ni-substrate interface is modulated as a result of induced stress by applying electrical voltage to the PMN-PT substrate.

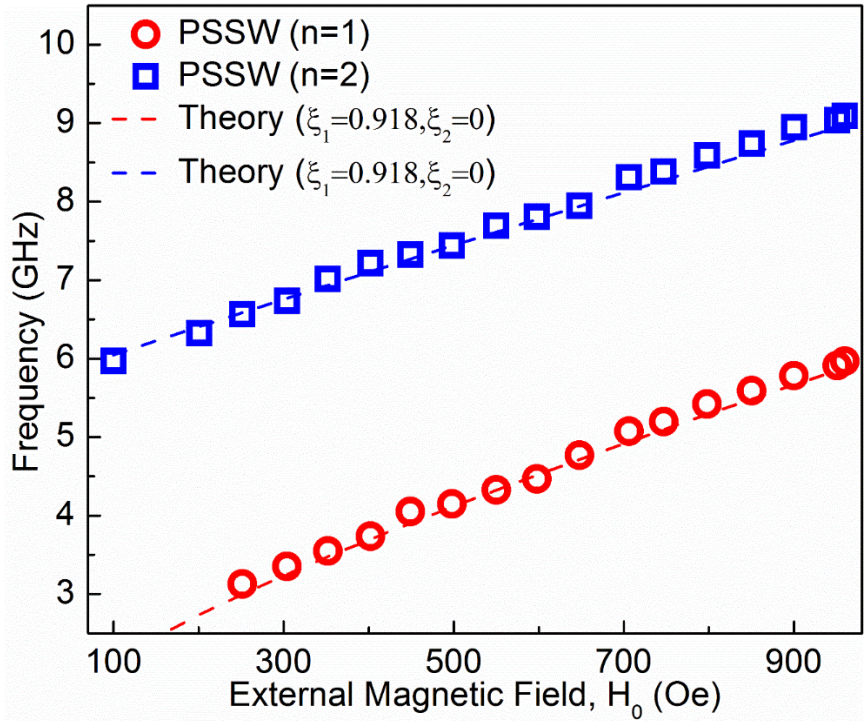

**Figure 4:** Frequency of the first and second PSSW modes as a function of the applied external magnetic field. The blue squares and red circles represent the frequencies of the PSSW modes associated with the two distinct peaks observed in BMS experiments. The blue and red dashed lines are theoretical fittings obtained assuming the partially pinning boundary condition at Ni-substrate and unpinned boundary condition at Ni-air interfaces, respectively.

The data presented in Figures 3 and 4 were obtained at zero voltage applied across the PMN-PT substrate. The next set of experiments was aimed to demonstrate the effect of the stress-mediated coupling between the piezoelectric and magnetostrictive layers on the spectral position of the confined modes. An electric field applied across the PMN-PT layer produces stress, which, in turn, affects the magnetic properties of the magnetostrictive Ni layer. The measured BMS spectra at several applied voltages and at constant external magnetic field $H_{eff} = 440$ Oe are shown in Figure 5. As one can see, the spectral positions of the peaks change with increasing electric field bias. Figure 6 shows the spectral position of each individual peak, observed in Figure 5, as a function of the applied electrical bias, at a constant external magnetic field $H_{eff} = 440$ Oe. The results indicate that the frequencies of the PSSW modes generally decrease with increasing electric field, although the decrease is non-monotonic. More importantly, the rates are different for the first and the second modes: 0.25 GHz / 0.4 MV/m for the first and 0.5 GHz / 0.4 MV/m for the second confined modes. Overall, the effect of applying the electric field is opposite to the one produced by the applied magnetic field.

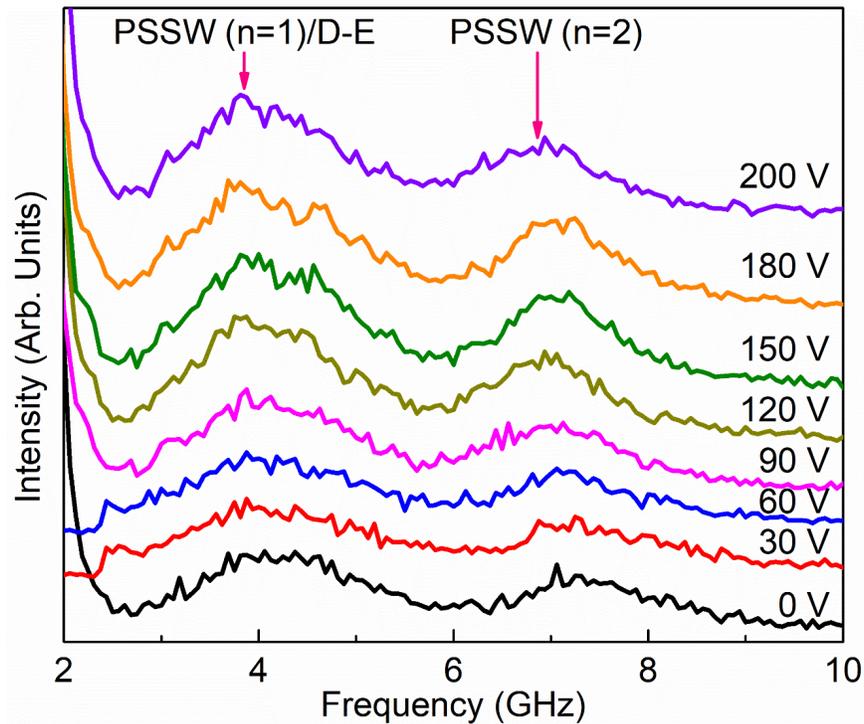

**Figure 5:** BMS spectra of the Ni thin film as a function of the applied external electric field to the piezoelectric PMN-PT substrate at a constant external magnetic field $H_{eff} = 440$ Oe. Application of the external electrical bias changes the frequency of the PSSW modes due to the induced stress and strain in the substrate and thin film.

Application of DC voltage bias produces a mechanical strain in PMN-PT due to its piezoelectric properties. This strain is transferred to the magneto-elastic Ni film and, in turn, changes its magnetic anisotropy[45], and corresponding conditions for the PSSW resonance modes. In particular, these changes reveal themselves in the appearance of an additional anisotropy field $H_a(V)$. This field, in turn, leads to a change in the frequency of the PSSW modes. It should be noted that although the frequency of both modes shifts down with the increasing electrical bias, the rate of the decrease is significantly larger for the second mode. This cannot be explained only by changes in the anisotropy field because based on Eq. (1), the sensitivity of the frequency to $H_a$ is approximately the same for both modes, and even a little smaller for second mode (see Supplementary Information). On the other hand, the sensitivity of the frequency to changes in the wave vector values is much larger for the second mode. Since the wave vector is completely defined by Eq. (3), its changes can take place in the case when the thickness of the film $d$ or/and pinning parameter $\xi$ change. For these reasons, we attribute the observed frequency shifting of PSSW to the applied DC voltage on the PMN-PT substrate not only to the changes in the magnetic properties of the thin film, but also to the thickness variations of the Ni film and the conditions for the pinning of the spin waves at Ni/PMN-PT interface.

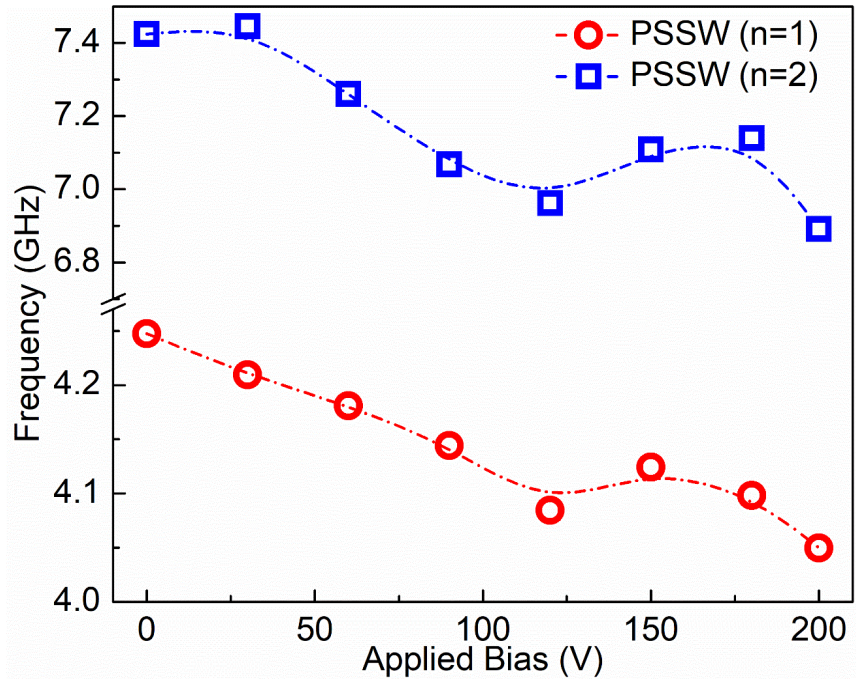

**Figure 6:** Frequency of the first (red circles) and second (blue squares) PSSW modes as a function of applied bias at a constant external magnetic field $H_{eff} = 440$ Oe. The blue squares and red circles show the experimental data from BMS experiments. Nota that the frequency of both modes decreases non-monotonically with the increasing electric field bias.

In summary, we used µ-BMS to investigate the spatially confined thermal magnons in the two-phase synthetic multiferroic structures consisting of a piezoelectric $[Pb(Mg_{1/3}Nb_{2/3})O_3]_{(1-x)} - [PbTiO_3]_x$ substrate and a Ni thin film. BMS spectra revealed two dominant peaks, which were attributed to the first and second PSSW modes, and described within the framework of the PSSW resonance with the asymmetrical boundary conditions at different interfaces of Ni film. Our data demonstrate the control of thermal magnons using the stress-mediated coupling in synthetic multiferroic structures with spatial confinement. An application of 0.4 MV/m electric field results in about 0.25 GHz and 0.5 GHz shifts for the first and the second confined magnon modes, respectively. The obtained results are important for understanding the thermal magnon spectrum in ferromagnetic films and development of the low-power spin-wave devices.

**Acknowledgements**

The work at UC Riverside was supported as part of the Spins and Heat in Nanoscale Electronic Systems (SHINES), an Energy Frontier Research Center funded by the U.S. Department of Energy, Office of Science, Basic Energy Sciences (BES) under Award # SC0012670. The work at UCLA was supported by the National Science Foundation under Cooperative Agreement Award EEC-1160504 for Center for Translational Applications of Nanoscale Multiferroic Systems (TANMS). The authors are indebted to Z. Barani for her help with the device schematic.

**Contributions**

A.A.B. and A.K. conceived the idea of the study. A.A.B. coordinated the project and contributed to the experimental and theoretical data analysis; F.K. and M.B. conducted µ-BMS and MOKE experiments and data analysis; H.C. carried out the sample characterization; G.K. prepared the samples; F.K. led the manuscript preparation. All authors contributed to writing and editing of the manuscript.